\documentclass[preprint,aps,prc,nofootinbib,showpacs]{revtex4}

\usepackage{epsfig}
\usepackage{graphicx}
\newcommand\ba{\begin{eqnarray}}
\newcommand\ea{\end{eqnarray}}
\newcommand\nn{\nonumber}
\newcommand{\be}{\begin{equation}}
\newcommand{\ee}{\end{equation}}

\newcommand{\bas}{\begin{eqnarray*}}
\newcommand{\eas}{\end{eqnarray*}}

\begin{document}

\title{Cross section and polarization observables for the reaction
$e^++e^-\to a_1(1260)+\pi $ }

\author{E. Tomasi--Gustafsson}
\affiliation{\it IRFU/Service de Physique Nucl\'eaire, CEA/Saclay, 91191 Gif-sur-Yvette, 
France}

\author{G. I. Gakh}
\affiliation{ National Science Centre "Kharkov Institute of Physics and Technology",\\ 61108 Akademicheskaya 1, Kharkov,
Ukraine } 

\author{C. Adamu\v s\v c\'in}
\affiliation{Institute of Physics, Slovak Academy of Sciences, Bratislava, Slovakia}

\date{\today}

\vspace{0.5cm}
\begin{abstract}
A model independent formalism for the electron positron annihilation reaction $e^++e^-\to a_1(1260)+\pi $ has been derived. The differential and total cross sections and the elements of the spin--density matrix of the $a_1$-meson were calculated in terms of the electromagnetic form factors of the corresponding  $\gamma^*a_1\pi $ current. Simple models of $a_1$ form factors have been fitted to the available cross section data and they allow to give numerical predictions for the different observables. 

\end{abstract}

\maketitle
 
\section{Introduction}

The electron positron annihilation into hadrons constitutes an important source of information on the internal structure of the mesons: the light quarks and their interactions as well as the spectroscopy of their bound states. The 
experimental data about these reactions in the low--energy region are also relevant to the determination of the strong interaction contribution to the anomalous magnetic moment of the muon, to the test of standard model predictions for the hadronic tau--lepton decay, which is related by the conservation of vector currents.

Recently, the construction of the new detectors with a large solid angle, 
which can operate at new colliders with high luminosity, 
opened new possibilities for the investigation of the reactions $e^++e^-\to$ multihadrons \cite{AKH99}. Not only the statistic is highly increased, but also the possibility to detect charged as well as neutral pions allows to draw conclusions on the nature of the intermediate states. 

In the energy region $1\leq W\leq 2.5$~ GeV ($W$ is the total energy of the colliding beams) the process of
four pion production is one of the dominant processes of the reaction $e^++e^-\to $hadrons. Its cross section is larger than $2\pi$ production and comparable to  $e^+ +e^-\to \mu^+ + \mu^-$. 

The process of  $e^+e^-$ annihilation into four pions was firstly detected in Frascati \cite{Ba70} and later on in Novosibirsk \cite{KU72}. Through a simultaneous analysis of the differential distributions in two final channels: $2\pi^+ 2\pi^-$ and $\pi^+\pi^- 2\pi^0$, it was shown in \cite{AKH99} that the reaction predominantly occurs through the $a_1(1260)\pi $ and $\omega\pi^0$ intermediate states in the energy range 1.05--1.38 GeV. It was also found that the relative fraction of the $a_1(1260)\pi $ state increases with the 
beam energy. The measurement of the $e^+e^-\to \pi^+\pi^-\pi^+\pi^-$ 
cross section was extended to lower energies. Data obtained with larger statistical and systematic precision \cite{AKH04} confirmed that the dominant  production mechanism is consistent with the  $a_1(1260)\pi $ intermediate state. 

The process of the multihadron production at large energies was also investigated with the BABAR detector at 
the PEP--II asymmetric electron--positron storage ring using the initial--state radiation \cite{Au05}. In 
particular, the cross section for the process $e^+e^-\to \pi^+\pi^-\pi^+\pi^-$ was measured for center--of--mass (CMS)  
energies from 0.6 to 4.5 GeV, providing evidence of a  
resonant structure, with preferred quasi--two--body production 
of $a_1(1260)\pi $. A detailed understanding of the four--pion final state requires also information from final states such as $\pi^+\pi^-\pi^0\pi^0$, to which the $\rho^+\rho^-$ intermediate state can contribute. A summary of the hadronic cross section 
measurements performed with BABAR via radiative return is given in Ref. \cite{De06}. 

In this paper we consider the reaction 
\be
e^-(k_1)+e^+(k_2)\to a_1(p_1)+\pi (p_2),
\label{eq:eq1}
\ee
where $a_1$ is the axial--vector meson $a_1(1260)$ with the following quantum numbers $I^G(J^{PC})=1^- (1^{++})$. The notation for the particle four-momenta is given in brackets.

The determination of the pseudoscalar--meson FFs as pions and kaons requires only cross section measurements. They have been extensively studied both in the space--like and time--like regions (see, for 
example, \cite{BV}). Light vector mesons are less known, because their experimental determination is more difficult, due to 
their short lifetimes. However, the $t$ dependence of the cross section for diffractive vector--meson 
electroproduction gives (model--dependent) information on the charge radius, and radiative decays, such as 
$\rho^+\to \pi^+\pi^0\gamma$, allow to obtain their magnetic moment.

From the theoretical point of view, the processes of the vector-- and axial--mesons production in the 
electron--positron annihilation were considered in a number of papers. The predictions for the differential and total 
cross sections of the reaction $e^+e^-\to a_1^{\pm}+\pi^{\mp}$ was given in Ref. \cite{VP71} in the framework  
of the hard--pion current algebra models. 

It was shown that the reaction cross 
section alone could be, in principle, discriminative toward models. Using VMD model, the authors of 
Ref. \cite{KUW71} investigated the reaction $e^+e^-\to mesons$ assuming two--body (or quasi--two--body) final states, as $a_1(1260)\pi $ and $\rho^+\rho^-$. All FFs were taken equal to unity. Estimations of the cross sections of 
the processes $e^+e^-\to 3\pi$, $4\pi$ were also obtained, using the VMD model, in Ref. \cite{AK72}. 

It appears that the magnitude and energy dependence of the cross section alone can not constitute a decisive test on the validity of VMD models. A recent  discussion can be found in Ref. \cite{Li07}, where existing models \cite{ESK94,PB96,Ac05} have been phenomenologically modified, including parameters to be fitted on the data. 
A good description of the cross section is obtained assuming $a_1\pi$ intermediate state, in addition to $\rho$ and $\pi$, and including higher $\rho$ resonances for energies over 1 GeV. 

Due to the conservation of vector current the cross section of the $e^+e^-\to 4\pi$ process can be related 
to the probability of the $\tau\to 4\pi\nu_{\tau}$ decay. Therefore, all realistic models describing the first 
process, should also be applicable to the description of the latter one. It was found \cite{BEMR99} that the 
assumption of the $a_1(1260)\pi $ dominance is in qualitative agreement with all available data. The 
free--parameter investigation of the branching ratios and distribution functions of the four particle decay  of 
$\tau\to \rho\pi\pi\nu $, in terms of the effective chiral theory of mesons, is consistent with the data \cite{Li98}.
The theory predicted the $a_1$ dominance in these four particle decay of the tau--lepton.

The purpose of this paper is to calculate the differential (and total) cross sections and the elements of the spin--density matrix of the $a_1$-meson in terms of the electromagnetic form factors (FFs) of the corresponding  $\gamma^*a_1\pi $ current. A model independent formalism, derived in \cite{GTG06} for spin one particles, and applied to the process $e^+ +e^-\to \rho^++\rho^-$ in \cite{Ad07}, allows to express the experimental observables (differential cross section, polarization observables, elements of the density matrix..) in terms of hadron FFs. In annihilation reactions, these FFs should be known, or extrapolated from the space-like region into the time-like (TL) region, on the basis of analytical arguments.
%%%%%%%%%%%%%%%%%%%%%%%%%%%
\section{Formalism}
%%%%%%%%%%%%%%%%%%%%%%%%%%%

The following derivation is based on the one-photon exchange mechanism. In principle, at large $q^2$, one should take into account the two--photon--exchange contribution, as it was suggested a few decades ago \cite{TPE}.  In case of spin one particles, model independent properties of the two--photon--exchange 
contribution in elastic electron--deuteron scattering have been discussed in Refs. \cite{Re99,Ga06}. However no experimental evidence has been found, up to now, on the presence of two--photon--exchange in the scattering \cite{ETG04} as well as in the annihilation channels \cite{ETG08}, and we do not include such contribution which makes the formalism very complicated. 

%%%%%%%%%%%%%%%%%%%%%%%%%%%%%%%%%%%%%%%%%%%%%%%%%
\subsection{The spin structure of the matrix element}
%%%%%%%%%%%%%%%%%%%%%%%%%%%%%%%%%%%%%%%%%%%%%%%%%%%%%%

In the one-photon approximation, the differential cross section of the reaction (\ref{eq:eq1}) in terms of the hadronic, $W_{\mu\nu}$, and leptonic, $L_{\mu\nu}$, tensors, neglecting the electron mass, is written as
\be\label{eq:eq2}
\frac{d\sigma}{d\Omega } = \frac{\alpha^2}{q^6}\frac{p}{2W}L_{\mu\nu}W_{\mu\nu},
\ee
where $\alpha=1/137$ is the electromagnetic constant, $p=\sqrt{(q^2+m^2-M^2)^2-4m^2q^2}/2W$ is the final--particle 
momentum in the reaction CMS, m and M are the masses of the pion and of the $a_1$ meson, respectively. The four 
momentum of the virtual photon is $q=k_1+k_2=p_1+p_2$, with $q^2=W^2$, and $W$ is the total energy of the initial beams (note that 
the cross section is not averaged over the spins of the initial beams).

The leptonic tensor (for the case of longitudinally polarized electron beam) is
\be
L_{\mu\nu}=-q^2g_{\mu\nu}+2(k_{1\mu}k_{2\nu}+k_{2\mu}k_{1\nu}) +
2i\lambda \varepsilon_{\mu\nu\sigma\rho}k_{1\sigma}k_{2\rho}\ , 
\label{eq:eq3}
\ee
where $\lambda$ is the degree of the electron beam polarization (further we assume that the electron beam is completely 
polarized and consequently $\lambda=1$). 

The hadronic tensor can be expressed via the electromagnetic current $J_{\mu}$, describing the transitions $\gamma ^*\rightarrow \pi a_1 $, as follows
\be
W_{\mu\nu} = J_{\mu}J^*_{\nu}.
\label{eq:eq4}
\ee
The hadron tensor $W_{\mu\nu}$ can be expressed in terms of FFs of the $\gamma ^*\rightarrow \pi a_1 $ transition, using the explicit form of the electromagnetic current 
$J_{\mu}$. The spin--density matrix of the $a_1-$ meson is composed of three terms, corresponding to unpolarized, vector and tensor polarized meson:
\be
\rho_{\mu\nu}=-\left(g_{\mu\nu}-\frac{p_{1\mu}
p_{1\nu}}{M^2}\right) +\frac{i}{2M}
 \varepsilon_{\mu\nu\rho\sigma}s_{\rho} p_{1\sigma}
+3Q_{\mu\nu}.
\label{eq:eq5} 
\ee 
Here $s_{\mu}$ and $Q_{\mu\nu}$ are the $a_1-$ meson polarization four vector and quadrupole tensor, 
respectively. The four vector of the $a_1-$ meson vector polarization $s_{\mu}$ and the $a_1-$
meson quadrupole--polarization tensor $Q_{\mu\nu }$ satisfy the following conditions:
 $$s^2=-1,~ sp_1=0,~Q_{\mu\nu}=Q_{\nu\mu}, \ \ Q_{\mu\mu}=0, ~
p_{1\mu}Q_{\mu\nu}=0\ .$$ 

Taking into account Eqs. (\ref{eq:eq4}) and (\ref{eq:eq5}), the hadronic tensor in the general case can be written as the sum of three terms
\be
W_{\mu\nu}=W_{\mu\nu}(0)+W_{\mu\nu}(V)+W_{\mu\nu}(T),
\label{eq:eq9}
\ee
where $W_{\mu\nu}(0)$ corresponds to the case of unpolarized particles in the final state and $W_{\mu\nu}(V)$$(W_{\mu\nu}(T))$ 
corresponds to the case of the  vector (tensor) polarized $a_1-$ meson. 

These expressions are general, for any spin one particle in the final state. Let us consider more particularly, the reaction $e^++e^-\rightarrow  \pi+a_1$ which has been shown to be the main contribution 
to the $4\pi $ final state.

The electromagnetic current of the $\gamma^* \to \pi a_1$ transition is described by two FFs. Assuming 
the P-- and C--invariance of the hadron electromagnetic interaction this current can be written as \cite{AK72}
\be\label{eq:eq5a}
J_{\mu}=f_1(q^2)(q^2U_{\mu}^*-q\cdot U^*q_{\mu})+f_2(q^2)(q\cdot p_2 U_{\mu}^*-q\cdot U^*p_{2\mu}), 
\ee
where $U_{\mu}$ is the polarization four-vector describing the spin one $a_1$--meson, and $f_i(q^2), (i=1, 2)$ 
are the electromagnetic FFs describing the $\gamma^* \to \pi a_1$ transition (note that we singled out 
explicitly the electron charge $e$ from the expression for the electromagnetic current). The FFs $f_i(q^2)$ 
are complex functions of the variable $q^2$ in the region of the TL momentum transfer ($q^2>0$). 

In case of real photon, $f_1$ does not contribute, and the value $f_2(0)$ can be obtained from the experimental data on the decay width $\Gamma (a_1\to \pi\gamma).$ The expression of the radiative decay width $ \Gamma$ in 
the axial--vector meson rest frame is
\be
\Gamma =\frac{1}{\pi}\frac{\omega}{24M^2}|{\cal M}|^2, 
\label{eq:eq9a}
\ee
where $\omega $ is the photon momentum in the axial--vector meson rest frame, $\omega =(M^2-m^2)/2M,$ and 
the matrix element of the radiative decay is written as: 
${\cal M}=ee_{\mu}J_{\mu}$, where $e_{\mu}$ is the photon 
polarization four--vector. So, the value of FF $f_2$ at point $q^2=0$ can be obtained from the following expression
\be
f_2^2(0) =12\frac{\Gamma M^3}{\alpha (M^2-m^2)^3}. 
\label{eq:eq10}
\ee 
Another estimation of these FFs was obtained in Ref. \cite{AK72} where the processes $e^+e^-\to 3\pi$, $4\pi $ were 
studied. The spin structure of the matrix element of the $a_1\rho\pi $ transition is similar to the $\gamma^*a_1\pi $ transition. The two coupling constants describing the $a_1\rho\pi $ transition were 
estimated by using the method of hard pions and knowing the width of the $a_1$--meson (taken about 130 MeV) \cite{AK72}. The corresponding values: $g^2/4\pi \approx 1.3$ and $\beta =1.6$ were obtained. The following relations hold:
$$ef_1(q^2=m_{\rho}^2)=\frac{g}{m_{\rho}}, \ \ 
ef_2(q^2=m_{\rho}^2)=\frac{g}{m_{\rho}}(\beta -1). $$

The explicit expressions for the 
contributions to the hadronic tensor are:

\noindent\underline{ - Unpolarized term $W_{\mu\nu}(0)$:}

$$W_{\mu\nu}(0)=W_1(q^2)\tilde{g}_{\mu\nu}+\frac{W_2(q^2)}{M^2}
\tilde{p}_{1\mu}\tilde{p}_{1\nu} \ , \
\tilde{g}_{\mu\nu}=g_{\mu\nu}-\frac{q_{\mu}q_{\nu}}{q^2}\ , \ \
\tilde{p}_{1\mu}=p_{1\mu}-\frac{p_1q}{q^2}q_{\mu} \ , $$
where
\be
W_1(q^2)=-|q^2f_1+\frac{1}{2}(q^2-M^2+m^2)f_2|^2, \ \ 
W_2(q^2)=q^2\left (q^2|f_1+f_2|^2-M^2|f_2|^2\right ),
\label{eq:eq10a}
\ee

\noindent\underline{ - Term for vector polarization $W_{\mu\nu}(V)$:}
\ba
W_{\mu\nu}(V)&=&\frac{i}{M}V_1(q^2)\varepsilon_{\mu\nu\sigma\rho} s_{\sigma}q_{\rho}+
\frac{i}{M^3}V_2(q^2)[\tilde{p}_{1\mu}
\varepsilon_{\nu\alpha\sigma\rho} s_{\alpha}q_{\sigma}p_{1\rho}-\tilde{p}_{1\nu}
\varepsilon_{\mu\alpha\sigma\rho} s_{\alpha}q_{\sigma}p_{1\rho}]+ \nn\\
&&+\frac{1}{M^3}V_3(q^2)[\tilde{p}_{1\mu}
\varepsilon_{\nu\alpha\sigma\rho} s_{\alpha}q_{\sigma}p_{1\rho}+\tilde{p}_{1\nu}\varepsilon_{\mu\alpha\sigma\rho} s_{\alpha}q_{\sigma}p_{1\rho}],\nn\\ V_1(q^2)&=&-M^2(q^2+M^2-m^2)^{-1}|q^2f_1+\frac{1}{2}(q^2-M^2+m^2)f_2|^2, \nn\\
V_2(q^2)&=&-M^2q^2(q^2+M^2-m^2)^{-1}\left [q^2|f_1|^2+\frac{1}{2}(q^2-M^2+m^2)|f_2|^2+ \right .\nn\\
&&
\left .\frac{1}{2}(3q^2-M^2+m^2)\mathrm{Re}f_1f_2^*\right ], \nn\\
V_3(q^2)&=&-\frac{1}{2}M^2q^2\mathrm{Im}f_1f_2^*, 
\label{eq:eq11}
\ea
\noindent\underline{ - Term for tensor polarization $W_{\mu\nu}(T)$:}

\ba
W_{\mu\nu}(T)&=&T_1(q^2)\bar Q\tilde{g}_{\mu\nu}+T_2(q^2)\frac{\bar Q}{M^2}
\tilde{p}_{1\mu}\tilde{p}_{1\nu}+T_3(q^2)(\tilde{p}_{1\mu}\widetilde{Q}_{
\nu}+\tilde{p}_{1\nu}\widetilde{Q}_{\mu})+\nn\\
&&T_4(q^2)\widetilde{Q}_{\mu\nu}\ +
iT_5(q^2)(\tilde{p}_{1\mu}\widetilde{Q}_{\nu}-\tilde{p}_{1\nu}\widetilde{Q}_{\mu}), 
\label{eq:eq12}
\ea
where
\ba
&\widetilde{Q}_{\mu}&=Q_{\mu\nu}q_{\nu}-\frac{q_{\mu}}{q^2}\bar
{Q},~ \widetilde{Q}_{\mu}q_{\mu}=0,~ 
\widetilde{Q}_{\mu\nu}=
Q_{\mu\nu}+\frac{q_{\mu}q_{\nu}}{q^4}\bar Q-
\frac{q_{\nu}q_{\alpha}}{q^2}Q_{\mu\alpha}-
\frac{q_{\mu}q_{\alpha}}{q^2}Q_{\nu\alpha},\nn\\
&\widetilde{Q}_{\mu\nu}q_{\nu} &= 0, \ \bar Q=Q_{\alpha\beta}q_{\alpha}q_{\beta},\nn\\
&T_1(q^2)&=0,~
T_2(q^2)=3M^2|f_2|^2, \ 
T_3(q^2)=\frac{3}{2}(q^2-M^2+m^2)|f_2|^2+3q^2\mathrm{Re}f_1f_2^*,\nn\\
&T_4(q^2)&=3|q^2f_1+\frac{1}{2}(q^2-M^2+m^2)f_2|^2, \
T_5(q^2)=-3q^2\mathrm{Im}f_1f_2^*. 
\label{eq:eq13}
\ea
%%%%%%%%%%%%%%%%%%%%%%%%%%%%%%%%%%%%%%%%%%%%%%%%%%%%%%%%%%%%%%%%%%%%%%%
\subsection{Expressions for the observables}
%%%%%%%%%%%%%%%%%%%%%%%%%%%%%%%%%%%%%%%%%%%%%%%%%%%%%%%%%%%%%%%%%%%%%%%

Using the definitions of the cross section (\ref{eq:eq2}), of the leptonic (\ref{eq:eq3}) and
hadronic (\ref{eq:eq9}) tensors, one can derive the expression for the unpolarized
differential cross section in terms of the structure functions $W_{1,2}$
(after averaging over the spins of the initial particles)
\be
\frac{d\sigma^{un}}{d\Omega }=\frac{\alpha^2}{2q^4}\frac{p}{W} \left \{
-W_1(q^2)+\frac{1}{2}W_2(q^2) \left [\tau -1-\frac{(u-t)^2}{4M^2q^2}
+\frac{M^2-m^2}{4M^2q^2}(2q^2+M^2-m^2)\right ]\right \}, 
\label{eq:eq15}
\ee
where $\tau =q^2/(4M^2)$, $t=(k_1-p_1)^2$ and $u=(k_1-p_2)^2$.
In the reaction CMS this expression can be written as
$$
\frac{d\sigma^{un}}{d\Omega }=\frac{\alpha^2}{2q^4}\frac{p}{W}(A+B\sin^2\theta ),
$$
\be
A=|q^2f_1+\frac{1}{2}(q^2-M^2+m^2)f_2|^2,
B=2\tau p^2[q^2|f_1+f_2|^2-M^2|f_2|^2], 
\label{eq:eq16}
\ee
where $\theta $ is the angle between the momenta of the axial--meson (${\vec p}$)
and of the electron beam (${\vec k}$). Integrating this expression with respect to the axial--meson angular
variables one obtains the following formula for the total cross section:
\ba 
\sigma_{tot}(e^+e^-\to \pi a_1)&=&
\frac{2\pi\alpha ^2}{3q^4}\frac{p}{W}\left [3|q^2f_1+\frac{1}{2}(q^2-M^2+m^2)f_2|^2+ \right .\nn \\
&& \left .+4\tau p^2[q^2|f_1+f_2|^2-M^2|f_2|^2 ]\right ]. 
\label{eq:eq16a}
\ea
Let us define an angular asymmetry, $R$, with respect to the
differential cross section, $\sigma_{\pi/2}$, measured at $\theta =\pi /2$, 
\be
\frac{d\sigma^{un}}{d\Omega }=\sigma_{\pi/2}(1+R\cos^2\theta ), R=-B/(A+B).  
\label{eq:i2}
\ee
As it was previously shown in the case of $e^+ +e^-\to d+\bar d$ \cite{GTG06}, this observable is very sensitive to the different underlying
assumptions on the axial--meson FFs and does not require polarization measurements.

The differential cross section in terms of FFs $f_i(q^2)$ contains not only the moduli of these 
FFs but also their interference (\ref{eq:eq16}). One can choose a linear combinations of these FFs, in such a way that the unpolarized 
differential cross section will contain only moduli. Let us introduce new FFs $g_i(q^2)$ which 
related to the old ones as follows
\be
f_1=g_1+g_2, \ \   f_2=cg_1+dg_2,
\label{eq:i3}
\ee
where $c=-2q^2/(q^2-M^2+m^2),$ $d=(q^2+M^2-m^2)/(M^2-q^2+m^2).$ Then the structure functions $W_i$ describing 
the unpolarized part of the hadronic tensor can be written as
\be
W_1(q^2)=-\frac{4p^4q^4}{(q^2-M^2-m^2)^2}|g_2|^2,
\label{eq:i4}
\ee
\be
W_2(q^2)=4p^2q^4\left [\frac{q^2}{(q^2-M^2+m^2)^2}|g_1|^2-
\frac{M^2}{(q^2-M^2-m^2)^2}|g_2|^2\right ].
\label{eq:i5}
\ee

The cross section can be written, in the general case, as the sum of unpolarized and polarized terms, 
corresponding to the different polarization states and polarization directions of the incident and scattered particles:
\be\label{eq:eq21}
\displaystyle\frac{d\sigma}{d\Omega}=
\displaystyle\frac{d\sigma^{un}}{d\Omega}
\left [1+P_y+\lambda P_x+\lambda P_z+ P_{zz}R_{zz}+
P_{xz}R_{xz}+P_{xx}(R_{xx}-R_{yy}) +\lambda P_{yz}R_{yz}\right ],
\ee
where $P_i$, $P_{ij}$, and $R_{ij}$, $i,j=x,y,z$ are, respectively, the components of the vector, tensor polarization  and  of the quadrupole polarization tensor of the outgoing 
$a_1 $--meson $Q_{\mu\nu}$, in its rest system and $\displaystyle d\sigma^{un}/d\Omega$ is the unpolarized differential  cross section. $\lambda$ is the degree of longitudinal polarization of the electron beam. It is explicitly indicated, in order to stress 
that these specific polarization observables are induced by the beam polarization.

Let us consider the different polarization observables and give their expression in terms of the $\gamma^*\to a_1\pi $
transition FFs.

\begin{itemize}
\item 

The vector polarization of the outgoing axial--meson, $P_y$, which does not require polarization in the initial 
state is
\be\label{eq:eq17}
P_y=\frac{1}{8}\frac{\sqrt{\tau }}{\sigma_0}\left [(q^2+M^2-m^2)^2-4M^2q^2 \right ]\sin(2\theta )\mathrm{Im}f_1f_2^*,
\ee
where $\sigma_0=A+B\sin^2\theta .$ One can see that this polarization is determined by non--zero phase difference 
of the complex FFs $f_1$ and $f_2$.

\item
The axial--vector meson can be tensor polarized also in case of unpolarized initial beams. 

The components of the tensor polarization are
\ba
P_{xx}&=&-\frac{3}{4}\frac{1}{\sigma_0}\sin^2\theta 
\left |q^2f_1+\frac{1}{2}(q^2-M^2+m^2)f_2\right |^2, \nn \\
P_{xz}&=&\frac{3}{4}\frac{\sqrt{\tau}}{\sigma_0}\frac{\sin(2\theta )}{q^2}
\left \{2(q^2+M^2-m^2)\left |q^2f_1+\frac{1}{2}(q^2-M^2+m^2)f_2 \right |^2+\right . \nn \\
&&+\left [ (q^2+M^2-m^2)^2-4M^2q^2\right ]
\left .\left[ \frac{1}{2}(q^2-M^2+m^2)|f_2|^2+q^2\mathrm{Re}f_1f_2^*\right]\right \}, \nn \\
P_{zz}&=&\frac{3}{8}\frac{1}{\sigma_0}\frac{1}{M^2q^2}\left[(q^2+M^2-m^2)^2-4M^2q^2
\right ]
\left \{\left|q^2f_1+\frac{1}{2}(q^2-M^2+m^2)f_2\right|^2+ \right .\nn \\
&& +\frac{1}{2}\sin^2\theta \left [ -q^4|f_1|^2+q^2(q^2+3M^2-3m^2)\mathrm{Re}f_1f_2^*+\right .
\nn \\
&&\left . \left .+\left( 2q^2(q^2-M^2)-\frac{3}{4}(q^2-M^2+m^2)^2\right )
|f_2|^2\right]\right\}. 
\label{eq:eq22}
\ea
A possible non--zero phase difference between the axial--meson FFs leads to another T--odd polarization 
observable proportional to the $R_{yz}$ component of the axial--meson tensor polarization. It takes the form
\be
P_{yz}=-\frac{3}{2}\frac{\sqrt{\tau}}{\sigma_0}\left [(q^2+M^2-m^2)^2-4M^2q^2\right ]
\sin\theta \mathrm{Im} f_1f_2^*. 
\label{eq:eq23}
\ee
\item 
Let us consider now the case of a longitudinally polarized electron beam. The other two components of the 
axial--meson vector polarization ($P_x$, $P_z$) require the initial particle polarization and are
\ba
P_x&=&-\frac{1}{4}\frac{\sqrt{\tau }}{\sigma_0}\sin\theta 
\left \{2q^2(q^2+M^2-m^2)|f_1|^2+\left[(q^2-M^2)^2-m^4\right ]|f_2|^2+\right .
\nn \\
&& \left . +\left[q^2(3q^2-2M^2-2m^2)-(M^2-m^2)^2\right ]\mathrm{Re}f_1f_2^*\right\}, \nn \\
P_z&=&\frac{1}{2}\frac{1}{\sigma_0}\cos\theta \left|q^2f_1+\frac{1}{2}(q^2-M^2+m^2)f_2\right |^2.
\label{eq:eq24}
\ea
\end{itemize}
%%%%%%%%%%%%%%%%%%%%%%%%%%%%%%%%%%%%%%%%%%%%%%%%%%%%%%%%%%%%%%%
\section{Spin--density matrix of axial--meson}
%%%%%%%%%%%%%%%%%%%%%%%%%%%%%%%%%%%%%%%%%%%%%%%%%%%%%%%%%%%%%%%%
\hspace{0.7cm}

For unstable particles, the vector and tensor polarizations are directly related to the angular distribution of their decay products; one can show that the angular distribution can be expressed in terms of the spin--density matrix. 
Let us calculate the elements of the spin--density matrix of the axial--meson which is produced in the reaction 
$e^++e^-\to \pi +a_1$. The calculation is done in CMS of this reaction. 

In case of unpolarized initial lepton beams, the convolution of the lepton $L_{\mu\nu}$ and hadron $W_{\mu\nu}$ tensors can be written as 
\be
S^{un}=S_{\mu\nu}U_{\mu}U_{\nu}^*,
\label{eq:eq29}
\ee
where $U_{\mu}$ is the polarization four--vector of the detected axial--meson and the $S_{\mu\nu}$ tensor can be 
represented in the following general form
\be
S_{\mu\nu}=S_1g_{\mu\nu}+S_2q_{\mu}q_{\nu}+S_3k_{1\mu}k_{1\nu}+
S_4(k_{1\mu}q_{\nu}+q_{\mu}k_{1\nu})+iS_5(k_{1\mu}q_{\nu}-q_{\mu}k_{1\nu}).
\label{eq:eq30}
\ee
The functions $S_i (i=1-5)$  can be written in terms of the two transition FFs of the 
axial--meson. Their explicit form is
\ba
S_1&=&-q^2\left|q^2f_1+\frac{1}{2}(q^2-M^2+m^2)f_2\right |^2, \nn \\
S_2&=&\left [q^2(q^2-M^2)-\frac{1}{4}(q^2-M^2+m^2+2Wp\cos\theta )^2
\right ]|f_2|^2+  \nn \\
&&+q^2(q^2+M^2-m^2-2Wp\cos\theta )\mathrm{Re}f_1f_2^*,   \nn \\
S_{3}&=& -4\left |q^2f_1+\frac{1}{2}(q^2-M^2+m^2)f_2\right |^2, \nn \\
S_4&=&2q^4|f_1|^2+2q^2(q^2-M^2+m^2+Wp\cos\theta )^2\mathrm{Re}f_1f_2^*+ \nn \\
&&+\frac{1}{2}(q^2-M^2+m^2)(q^2-M^2+m^2+2Wp\cos\theta )|f_2|^2,  \nn \\
S_5&=&-2q^2Wp\cos\theta \mathrm{Im}f_1f_2^*.
\label{eq:eq31}
\ea
The T--odd structure function $S_5$ is not zero here since the transition FFs of the axial--meson are complex functions.

The elements of the spin--density matrix of the axial--meson are defined as
\be
S\rho_{mm^{\prime}}=S_{\mu\nu}U_{\mu}^{(m)}U_{\nu}^{(m^{\prime})*},~
S=S_{\mu\nu} \left (-g_{\mu\nu}+\frac{p_{1\mu}p_{1\nu}}{M^2}\right ),
\label{eq:eqs4}
\ee
where $S=2q^2(A+B\sin^2\theta )$, and $U_{\mu}^{(m)}$ is the polarization
four--vector of the axial--meson with definite $(m=0, \pm 1)$ projection
on the $z$ axis. In our case it is directed along the axial--meson momentum
and thus $U_{\mu}^{(m)}$ are the polarization vectors with definite helicity.

The elements of the spin--density matrix of the axial--meson are
\ba
\rho_{++}&=&\rho_{--}=\frac{q^2}{2S}(1+\cos^2\theta )\left |q^2f_1+\frac{1}{2}(q^2-M^2+m^2)f_2\right |^2, \nn \\
\rho_{00}&=&\frac{q^4}{M^2S}\sin^2\theta \left \{\frac{1}{4}(q^2+M^2-m^2)^2|f_1|^2+
\left[2p^2q^2+M^2(q^2-M^2+m^2)\right]\mathrm{Re}f_1f_2^*+ \right . \nn \\
&& \left .+(m^2M^2+p^2q^2)|f_2|^2\right \}, \nn \\
\rho_{+-}&=&\rho_{-+}=\frac{q^2}{2S}\sin^2\theta
\left |q^2f_1+\frac{1}{2}(q^2-M^2+m^2)f_2\right|^2,
\nn \\
\rho_{+0}&=&-\frac{q^4}{S}\sqrt{\frac{\tau}{2}}\sin\theta \cos\theta
\left \{(q^2+M^2-m^2)|f_1|^2+(1-\frac{M^2-m^2}{q^2}) \left [(q^2+M^2-m^2)\mathrm{Re}f_1f_2^*+ \right . \right .\nn \\
&&\left . \left . +
\frac{1}{2}(q^2-M^2-m^2)|f_2|^2 \right]+2p^2f_2f_1^*\right \}, \nn \\
\rho_{-0}&=&-\rho_{+0},~ \rho_{0+}=\rho_{+0}^*,~
\rho_{0-}=\rho_{-0}^*. 
\label{eq:eqs4a}
\ea
The spin--density matrix is normalized as $Tr\rho =1$ or $\rho_{++}+
\rho_{--}+\rho_{00}=1.$ The element $\rho_{+0}$ is complex and the real and imaginary parts are written as:
\ba
\mathrm{Re}\rho_{+0}&=&-\frac{q^4}{S}\sqrt{\frac{\tau}{2}}\sin\theta \cos\theta
\left \{
\left(q^2+M^2-m^2\right )|f_1|^2+\frac{1}{2}\left(1-\frac{M^2-m^2}{q^2}\right )
\right .
\nn\\
&&
\left . \left(q^2-M^2-m^2\right )|f_2|^2+\frac{1}{2}\left [3q^2-2M^2-2m^2-\frac{(M^2-m^2)^2}{q^2}\right ]\mathrm{Re}f_1f_2^*\right \}, 
\nn\\
\mathrm{Im}\rho_{+0}&=&\frac{p^2q^4}{S}\sqrt{\frac{\tau}{2}}\sin2\theta
\mathrm{Im}f_1f_2^*. 
\label{eq:eqs5}
\ea
Let us consider the case when the electron beam is longitudinally polarized. Then the convolution 
of the spin--dependent part of the lepton and hadron tensors can be written as
\be
S(\lambda )=S_{\mu\nu}(\lambda )U_{\mu}U_{\nu}^*,
\label{eq:eqs5a}
\ee
where $\lambda $ is the degree of the electron beam polarization and the  $S_{\mu\nu}(\lambda )$ tensor can be written as
\be
S_{\mu\nu}(\lambda )=Q_1\epsilon_{\mu\nu\alpha\beta}k_{1\alpha}k_{2\beta}+
Q_2(q_{\mu}a_{\nu}-q_{\nu}a_{\mu})+Q_3(q_{\mu}a_{\nu}+q_{\nu}a_{\mu}),
\label{eq:eqs6}
\ee
where $a_{\mu}=\epsilon_{\mu\alpha\beta\gamma}p_{\alpha}k_{1\beta}k_{2\gamma}$,
$p=p_1-p_2$. The structure functions $Q_i (i=1-3)$ can be written in
terms of the two transition FFs of the 
axial--meson as
\ba 
Q_{1}&=&-2i\lambda \left |q^2f_1+\frac{1}{2}(q^2-M^2+m^2)f_2\right |^2, \nn \\
Q_{2}&=&-2i\lambda \mathrm{Re}\left [q^2f_1+\frac{1}{2}(q^2-M^2+m^2)f_2\right ]f_2^*,\nn \\
Q_3&=&2\lambda q^2\mathrm{Im}f_1f_2^*.
\label{eq:eqs7} 
\ea
The T--odd structure function $Q_3$ is not zero since FFs are complex functions in the TL region.

The elements of the spin--density matrix of the $a_1$-meson that depend
on the longitudinal polarization of the electron beam can be defined as
\be
S\rho_{mm^{\prime}}(\lambda)=
S_{\mu\nu}(\lambda )U_{\mu}^{(m)}U_{\nu}^{(m^{\prime})*},
\label{eq:eqs6a}
\ee
and they are expressed in terms of FFs as
\ba
\rho_{++}(\lambda )&=&-\rho_{--}(\lambda )=\frac{\lambda}{S}
q^2\cos\theta \left|q^2f_1+\frac{1}{2}(q^2-M^2+m^2)f_2\right |^2, \nn \\
\rho_{00}(\lambda )&=&\rho_{+-}(\lambda )=\rho_{-+}(\lambda )=0, \nn \\
\rho_{+0}(\lambda )&=&-\frac{\lambda}{S}\sqrt{\frac{\tau}{2}}
q^4\sin\theta 
\left \{(q^2+M^2-m^2)|f_1|^2+\right. \nn \\
&&
+\left . \left(1-\frac{M^2-m^2}{q^2}\right )
\left[(q^2+M^2-m^2)\mathrm{Re}f_1f_2^* 
+\frac{1}{2}(q^2-M^2-m^2)|f_2|^2\right ]+2p^2f_2f_1^*\right \}, \nn \\
\rho_{0+}(\lambda )&=&\rho_{+0}^*(\lambda ), ~
\rho_{-0}(\lambda )=\rho_{+0}(\lambda ), ~
\rho_{0-}(\lambda )=\rho_{-0}^*(\lambda ).
\label{eq:eqs7a}
\ea
The spin--density matrix element $\rho_{+0}(\lambda )$ is a complex quantity and its real and imaginary parts are
\ba
\mathrm{Re}\rho_{+0}(\lambda )&=&-\frac{\lambda}{S}\sqrt{\frac{\tau}{2}}
q^4\sin\theta 
\left \{(q^2+M^2-m^2)|f_1|^2+\frac{1}{2}\left (1-\frac{M^2-m^2}{q^2}\right )(q^2-M^2-m^2)|f_2|^2+
\right . \nn \\
&&\left .+\frac{1}{2}\left [(3q^2-2M^2-2m^2-\frac{(M^2-m^2)^2}{q^2}
\right ] \mathrm{Re}f_1f_2^*\right \}, \nn \\
\mathrm{Im}\rho_{+0}(\lambda )&=&\frac{\lambda}{S}\sqrt{2\tau}p^2q^4\sin\theta \mathrm{Im}f_1f_2^*.
\label{eq:eqs7b}
\ea
The axial--meson FFs are complex functions in the TL region. So, for the complete determination of FFs it is 
necessary to measure three quantities: two moduli of FFs and their phase difference. Therefore, the 
measurement of the unpolarized differential cross section does not allow to determine completely FFs. It is 
necessary to determine the spin--density matrix elements of the produced axial meson measuring the angular 
distribution of its decay products.

The measurement of the angular dependence of the unpolarized differential cross section allows to determine  
the structure functions $A$ and $B$. So, one can determine the following ratio
\ba
R_1&=&\frac{B}{A}=\frac{p^2}{2M^2}\left [1+2r\cos\alpha +\left(1-\frac{M^2}{q^2}\right)r^2\right ]
\nn \\
&&\left [1+\left(1+\frac{m^2-M^2}{q^2}\right)r\cos\alpha +\frac{1}{4}\left(1+\frac{m^2-M^2}{q^2}\right)^2r^2\right ]^{-1}, 
\label{eq:eqs8}
\ea
where $r=|f_2|/|f_1|$ and $\alpha $ is the relative phase of two complex FFs $f_1$ and $f_2$ defined as 
follows: $\alpha =\alpha_1-\alpha_2$, where $\alpha_1=\mathrm{arg}f_1$ and $\alpha_2=\mathrm{arg}f_2$. Thus, Eq. \ref{eq:eqs8} contains two unknown quantities: $\alpha $ and $r$. Another equation for the determination of these 
quantities can be obtained, for example, from the following ratio of the spin--density matrix elements of the produced axial meson
\ba
R_2&=&\frac{\rho_{00}}{\rho_{+-}}=8\tau \left \{\frac{1}{4}(q^2+M^2-m^2)^2+
 \left[2p^2q^2+M^2(q^2-M^2+m^2)\right]r\cos\alpha + \right .
\label{eq:eqs8a} \\
&&\left .+(m^2M^2+p^2q^2)r^2\right \}\left [q^4+q^2(q^2-M^2+m^2)r\cos\alpha +\frac{1}{4}(q^2-M^2+m^2)^2
r^2\right ]^{-1}. \nn
\ea
And, finally, the phase difference $\alpha $ can be determined by fixing the sign of $\sin\alpha $ from 
the measured spin--density matrix element $\mathrm{Im}\rho_{0+}$. Note that the complete determination of two complex
FFs $f_1$ and $f_2$ does not require the polarization of the initial beams. The measurement of the angular distribution of the unpolarized differential cross section allows to determine the structure functions $A$ an $B$ separately. Knowing the ratio $r$, one can determine the moduli $|f_1|$ and $|f_2|$.
%%%%%%%%%%%%%%%%%%%%%%%%%%%%%%%%%%%%%%%%%%%%%%%%%%%%%%%%%%%%%%%
\section{Numerical Results}
%%%%%%%%%%%%%%%%%%%%%%%%%%%%%%%%%%%%%%%%%%%%%%%%%%%%%%%%%%%%%%%

VMD inspired models have proved to be very successful in describing the structure of hadrons. Such models contain a small number of parameters, with transparent physical meaning, and can be analytically extended to the full region of momentum transfer squared. A monopole-like behavior reproduce quite well the existing experimental data on pion FFs, and satisfies pQCD asymptotic \cite{BH92}. In order to predict the behavior of polarization observables, we suggest a simple model for the $a_1$ transition FFs, in TL region.

We used a simple VMD-based parametrization saturated by  vector mesons. The  contribution of one vector meson is given by the Breit-Wigner form
\begin{equation}
 \Delta f_i = \frac{C_{v,i}M_v}{M_v^2-q^2+i M_v \Gamma_v}, \; i=1,2 ,
\end{equation}
where $M_v$ and $\Gamma_v$ are the mass and the width of a vector meson carrying the interaction. In general one should introduce all allowed vector mesons, but, as shown in Refs. \cite{Li07,Akh00}, the largest contribution to the cross section is given by the $\rho (770)$, $\rho '(1450)$, and, at higher energies,  $\rho "(1700)$. We consider only data for the total cross section at energies above the $a_1\pi$ kinematical threshold, which have been compiled from Refs. \cite{Bac80,Cor82,Esp80,Au05}. The experimental data \cite{Au05} show a clear contribution from $J/\psi$ around $\sqrt{s}=3.01$ GeV. We excluded the corresponding (four) data points from the fit. The final form of our parametrization of the two $\gamma^*\pi a_1$ transition form factors
\begin{equation}
 f_i=\frac{C_{\rho , i}M_{\rho}}{M_{\rho}^2-q^2+i M_{\rho}\Gamma_{\rho}}+\frac{C_{\rho' , i}M_{\rho'}}{M_{\rho'}^2-q^2+i M_{\rho'}\Gamma_{\rho' }}+\frac{C_{\rho" , i}M_{\rho" }}{M_{\rho" }^2-q^2+i M_{\rho" }\Gamma_{\rho" }}, \; i=1,2. \label{eq:vmdFFs}
\end{equation}
Such parametrization has, therefore, in total six parameters: three normalization constants, $C_{\rho , i}$, $C_{\rho' , i}$, $C_{\rho'' , i}$ for each FF.

\begin{figure}
	\centering
	\includegraphics[width=0.7\textwidth]{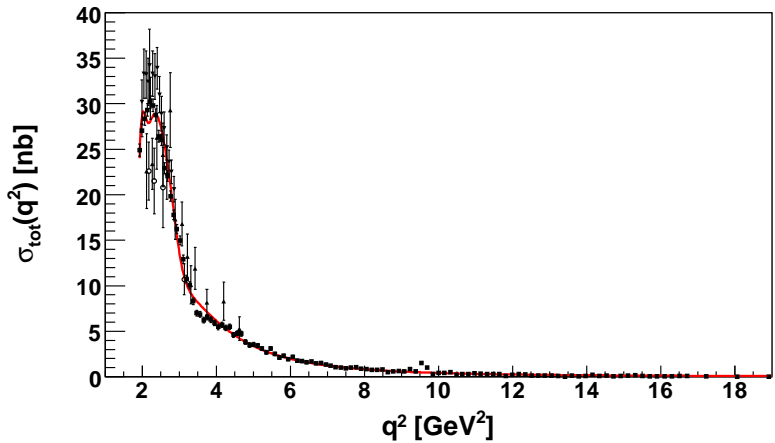}
        \caption{ (Color online) Total cross section data for the annihilation $e^+e^- \rightarrow \pi^+\pi^-\pi^+\pi^-$ from \protect\cite{Au05,Bac80,Esp80,Cor82}. The line represents the fit of the total cross section (\ref{eq:eq16a}) for the reaction $e^+ +e^- \rightarrow a_1 +\pi$ with VMD-based model of transition form factors (\ref{eq:vmdFFs}). }
      	\label{fig:1}
\end{figure}

\begin{figure}
	\includegraphics[width=1.\textwidth]{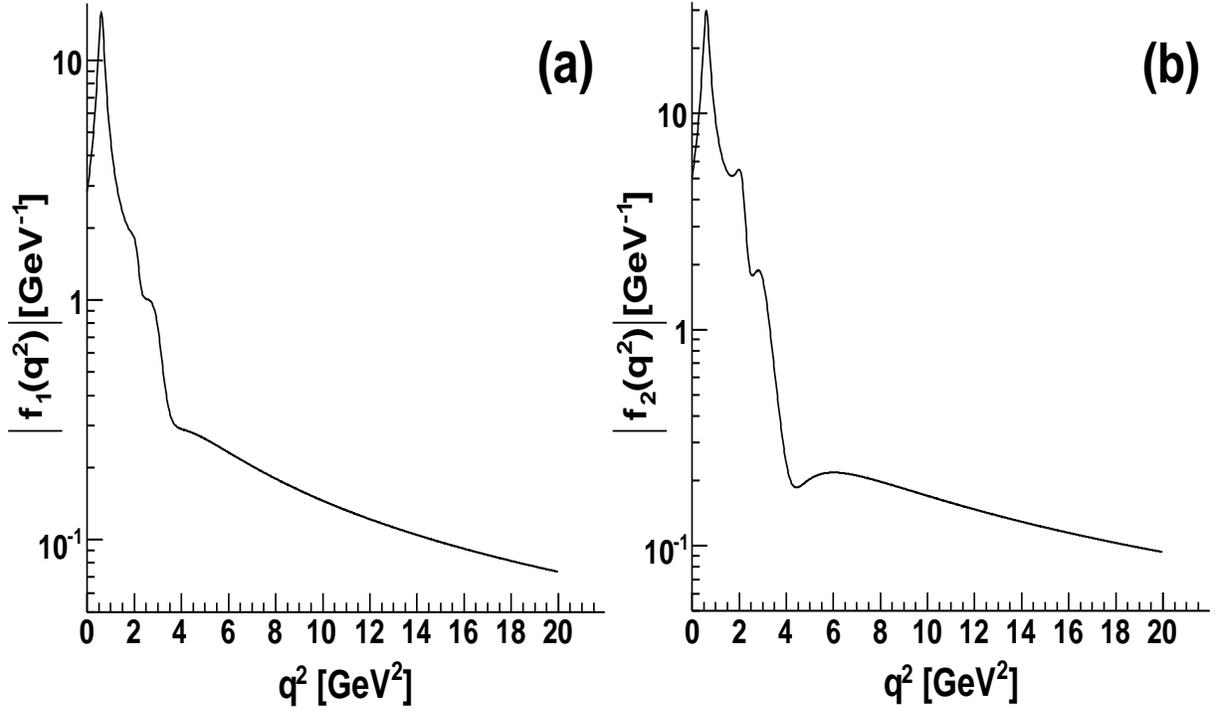}
        \caption{$q^2$ dependence of the $\gamma^*\pi a_1$ transition form factors $f_1$, $f_2$ (\ref{eq:vmdFFs}). }
      	\label{fig:1b}
\end{figure}

The result of the fit for the available data on the total cross section of the $e^+ e^- \rightarrow \pi^+\pi^-\pi^+\pi^-$ process from \protect\cite{Au05,Bac80,Esp80,Cor82} is shown in Fig. \ref{fig:1}, and corresponds to $\chi^2/ndf=455/154=2.9$. The different sets of data points are quite dispersed, especially in the region of the maximum. The resulting  normalization constants are given in Table \ref{table:1}. The $\gamma^*\pi a_1$ transition form factors (moduli) are presented in the Fig.\ref{fig:1b}. Peaks can be seen in correspondence with the masses of the chosen vector mesons. For $|f_2|$, one can see a bump around $q^2=4$ GeV$^2$, which results from the interference of the different terms.

Such fit assumes that the intermediate state $a_1\pi$ saturates the cross section \cite{AKH04}. If other intermediate channels contribute to this yield, and only a fraction of the cross section is due to the $a_1\pi$ intermediate state, then (assuming no dependence on $q^2$) the normalization parameters should be rescaled by the square root of that fraction.

\begin{figure}
	\includegraphics[width=1.0\textwidth]{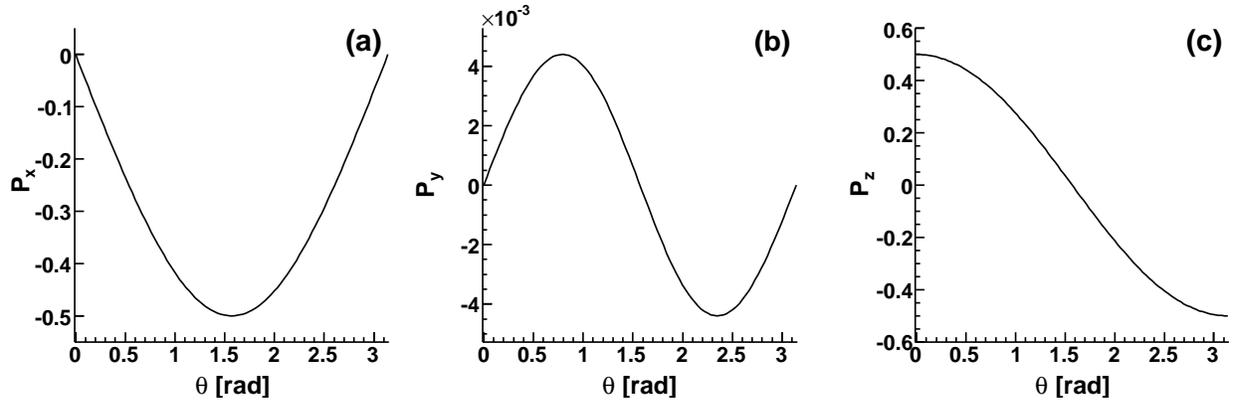}
        \caption{Angular dependence of the vector polarization observables $P_x$, $P_y$, $P_z$ at $q^2=2$ GeV$^2$, Eqs. (\protect\ref{eq:eq17},\protect\ref{eq:eq24}).}
      	\label{fig:2}
\end{figure}

\begin{figure}
	\centering
	\includegraphics[width=0.66\textwidth]{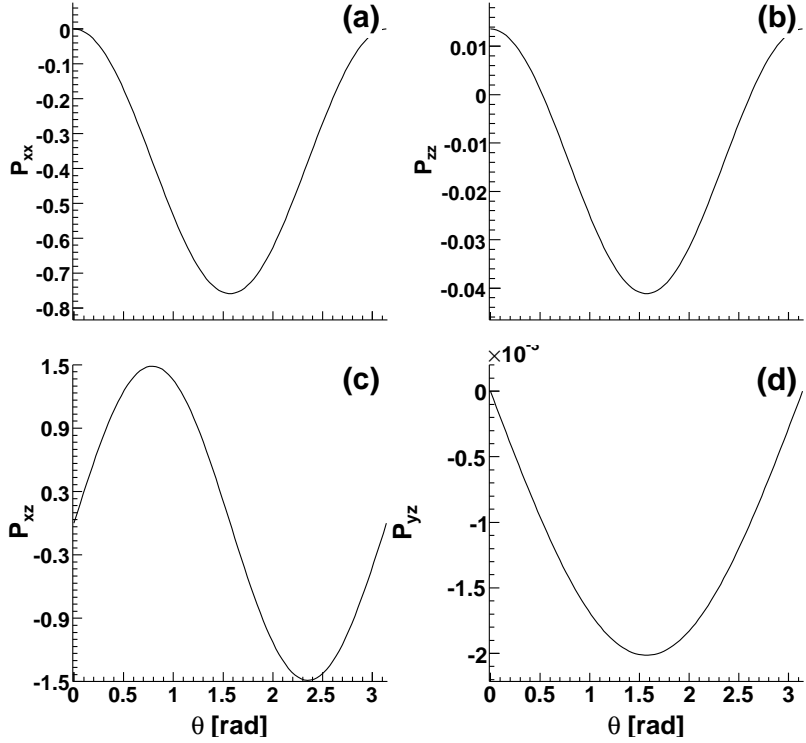}
        \caption{Angular dependence of the tensor polarization observables $P_{xx}$, $P_{xz}$, $P_{zz}$, $P_{yz}$ at $q^2=2$ GeV$^2$, Eqs.  (\protect\ref{eq:eq22},\protect\ref{eq:eq23}).}
      	\label{fig:3}
\end{figure}

\begin{figure}
	\includegraphics[width=1.0\textwidth]{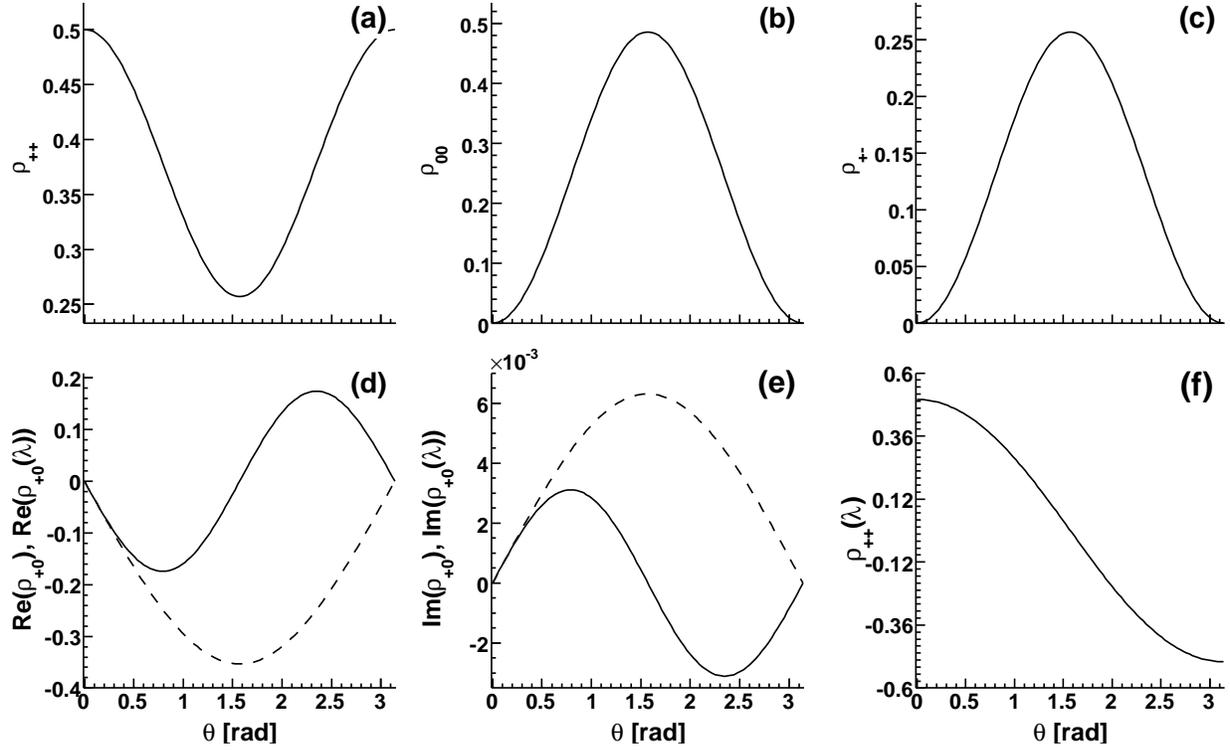}
        \caption{Angular dependence of the elements of the spin density matrix   $\rho_{++},\rho_{00},\rho_{+-},\rho_{+0}$ at $q^2=2$ GeV$^2$, in case of unpolarized collisison (solid lines) from Eqs. (\protect\ref{eq:eqs4a},\protect\ref{eq:eqs5}). In case of longitudinally polarized electrons the corresponding elements are shown also (dashed lines, (d),(e)) from Eqs. (\protect\ref{eq:eqs7b}) and (f) from Eq. (\protect\ref{eq:eqs7a}).}
      	\label{fig:4}
\end{figure}

\begin{figure}
	\includegraphics[width=1.0\textwidth]{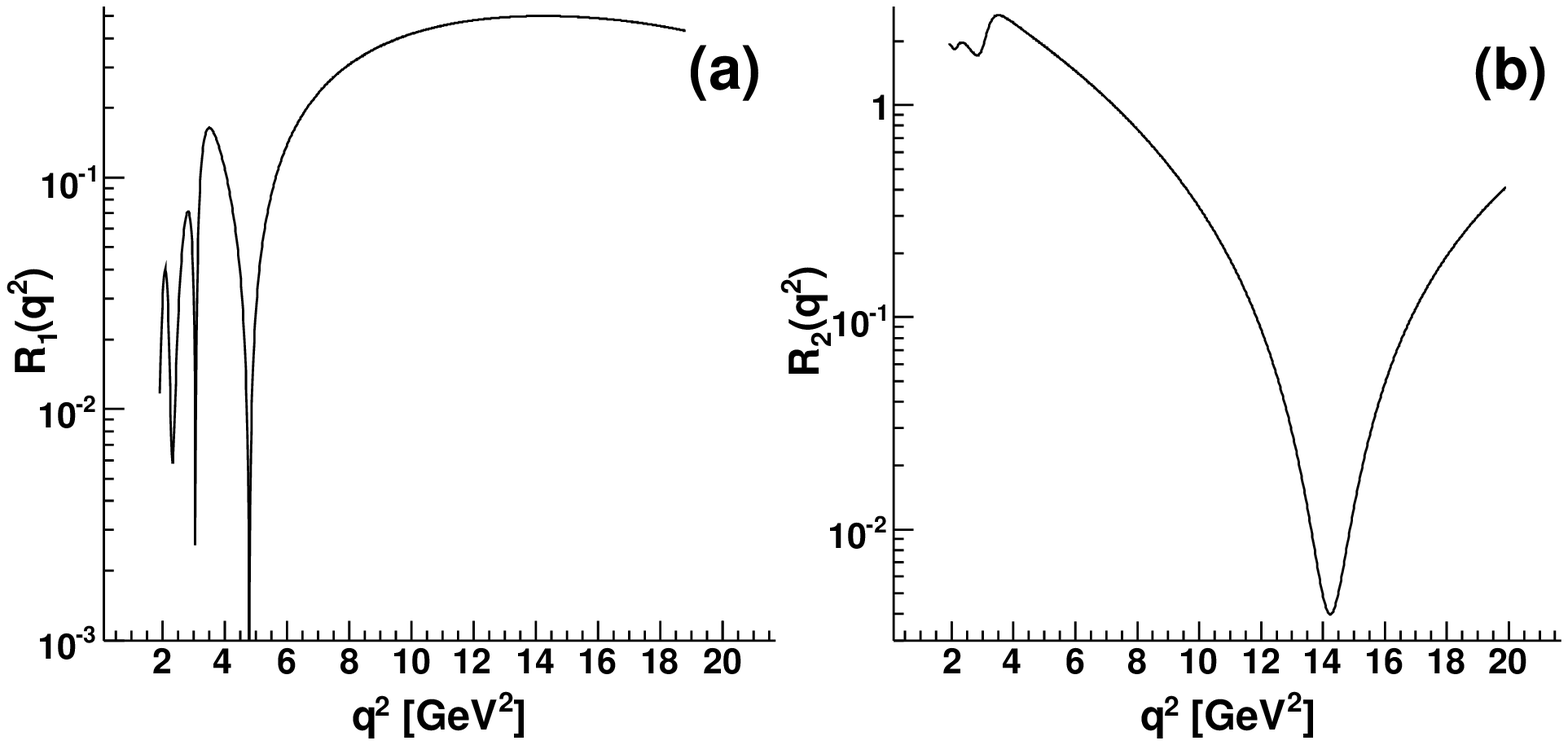}
        \caption{$q^2$ behavior of the ratios $R_1$, $R_2$, Eqs.  (\protect\ref{eq:eqs8},\protect\ref{eq:eqs8a}).}
      	\label{fig:5}
\end{figure}

Figs. \ref{fig:2}, \ref{fig:3} show the vector and tensor polarization observables. One should note here that the vector polarization of vector mesons 
can not be measured through their decays which are driven by strong and electromagnetic interaction with conservation of P-parity \cite{ETG02}. 
Fig. \ref{fig:4} shows the predictions for the density matrix elements. 
These quantities can be quite large and show a particular behavior, which can be experimentally verified.
Finally, Fig. \ref{fig:5} shows the predictions for the ratios  $R_1$, $R_2$ Eqs. (\ref{eq:eqs8},\ref{eq:eqs8a}).

\begin{table}
\centering
\begin{tabular}{|c|r|r|r|}
\hline\hline
 $i$ & $C_{\rho,i}~~~~$   & $C_{\rho' ,i}~~~~$&  $C_{\rho",i}~~~~$ \\
\hline
%1 & $1.82\pm 0.01$&$-0.13\pm 0.01$ &$-0.22\pm 0.02$ \\
%2 & $-3.45\pm 0.08$ &$0.82\pm 0.07$ &$0.66\pm 0.06$ \\
% *.775   * 1.459 * 1.720
1 & $2.35\pm 0.01$&$-0.089\pm 0.008$ &$-0.131\pm 0.009$ \\
2 & $-4.45\pm 0.11$ &$0.56\pm 0.05$ &$0.38\pm 0.03$ \\
\hline\hline
\end{tabular}
\caption{Optimal values of the normalization constants $C_{\rho,i},C_{\rho',i},C_{\rho",i}$ obtained by the fitting procedure.}
\label{table:1}
\end{table}

\section{Axial--meson decay to three pions}

For completeness, we derive the angular distribution for the decay of the $a_1$--meson to three pions
\begin{equation}\label{eq:eq3pi}
a_1^+(p)\to \pi^+(q_1)+\pi^-(q_2)+\pi^+(q_3),
\end{equation}
where the notation for the particle four--momenta is given in brackets. We
assume that this decay takes place through a quasi--two--body production
of $\rho\pi $ and $\sigma\pi $ where $\sigma $ (or $f_0(600)$) is a scalar
meson with the following quantum numbers $I^G (J^{PC})=0^+(0^{++}).$ The
$\rho $-- and $\sigma $--mesons are decaying predominantly to $\pi\pi $
final state.

The matrix element of this decay corresponding to the following mechanism
$a_1\to (\rho\pi +\sigma\pi )\to 3\pi $ can be written as
\begin{equation}\label{2}
M=aq_1\cdot U+bq_2\cdot U=U_{\mu}A_{\mu},
\end{equation}
where $U_{\mu}$ is the polarization four--vector describing the $a_1$--meson
and the functions $a$ and $b$ are (in the $a_1$--meson rest system):
\ba
a&=&G[R^{-1}(Q^2)-R^{-1}(q^2)]+\frac{1}{2}D^{-1}(q^2)[(F_2-F_1)(M^2+m^2-
2M\omega_1)-2MF_2\omega_2]+\nn\\
&&+\frac{1}{2}D^{-1}(Q^2)[(F_2-F_1)(M^2-m^2-
2M\omega_1)+2MF_1\omega_2], \nn\\
b&=&GR^{-1}(Q^2)-D^{-1}(q^2)[(F_2-F_1)(M\omega_1-m^2)+MF_1(M-\omega_1)]+ \nn\\
&&+\frac{1}{2}D^{-1}(Q^2)[(F_2-F_1)(m^2-M^2+2M\omega_2)-
2MF_1\omega_1], \nn\\
q^2&=&M^2+m^2-2M\omega_1, \ \  Q^2=m^2-M^2+2M(\omega_1+\omega_2).
\label{eq:eqab}
\ea
Here $\omega_1$ and $\omega_2$ are the energies of the positive and negative
pions. The expression for the matrix element takes into account the identity
of the two positive pions. For the intermediate $\rho $-- and $\sigma $--mesons
we use a standard Breit--Wigner form and
$$D(q^2)=q^2-m_{\rho}^2+i\Gamma_{\rho}m_{\rho}, \ \
R(q^2)=q^2-m_{\sigma}^2+i\Gamma_{\sigma}m_{\sigma}, $$
where $m_{\rho} (\Gamma_{\rho}) $ and $m_{\sigma} (\Gamma_{\sigma})$
are the masses (widths) of the $\rho $-- and $\sigma $--mesons, respectively.
The quantities $G$ and $F_{1,2}$ are defined as follows
$$G=g_{a\sigma\pi}g_{\sigma\pi\pi}, \ \  F_i=g_{\rho\pi\pi}\bar f_i,
\ \  i=1,2, $$
where $\bar f_i, \  i=1,2 $ are the coupling constants defining the
$a_1\to\rho\pi $ vertex.

The definition of the different coupling constants and their relation to the
corresponding decay width is given below.

\underline {The $\sigma\to\pi\pi $ decay.} The amplitude of this decay can be
written as: $M=g_{\sigma\pi\pi}\varphi_1^*\varphi_2^*\psi $, where $\varphi_i$
and $\psi $ are the wave functions of the pions and $\sigma $--meson,
respectively. Then the expression of the decay width in the $\sigma $--meson
rest frame is
\begin{equation}\label{4}
\Gamma (\sigma\to\pi\pi )=\frac{g^2_{\sigma\pi\pi}}{16\pi m^2_{\sigma}}
\sqrt{m^2_{\sigma}-4m^2}.
\end{equation}

\underline {The $\rho\to\pi\pi $ decay.} The matrix element describing this
decay can be written as: $M=g_{\rho\pi\pi}\varepsilon_{\mu}(q_1-q_2)_{\mu}
\varphi_1^*\varphi_2^* $, where $\varepsilon_{\mu}$ is the $\rho $--meson
polarization four--vector. $q_1(q_2)$ and $\varphi_1 (\varphi_2)$ are the
four--momentum and the wave function of the first (second) pion respectively. The expression
for the width of this decay in the $\rho $--meson rest frame is
\begin{equation}\label{5}
\Gamma (\rho\to\pi\pi )=\frac{g^2_{\rho\pi\pi}}{48\pi m^2_{\rho}}
(m^2_{\rho}-4m^2)^{3/2}.
\end{equation}

\underline {The $a_1\to\sigma\pi $ decay.} The matrix element in this case
is: $M=g_{a\sigma\pi}U_{\mu}(q_1-q_2)_{\mu}\varphi^*\psi^* $, where $U_{\mu}$
is the $a_1 $--meson polarization four--vector and $q_1(q_2)$ is the $\sigma
(\pi -)$--meson four--momentum. The width of this decay in the $a_1 $--meson
rest frame is
\begin{equation}\label{6}
\Gamma (a_1\to\sigma\pi )=\frac{g^2_{a\sigma\pi}}{48\pi M^5}
[(M^2+m^2-m^2_{\sigma})^2-4m^2M^2]^{3/2}.
\end{equation}

\underline {The $a_1\to\rho\pi $ decay.} The matrix element of this decay
is determined by two coupling constants and can be written as: $M=\bar f_1
(q^2U\cdot \varepsilon^*-q\cdot Uq\cdot \varepsilon^*)+\bar f_2(q\cdot p_2
U\cdot \varepsilon^*-q\cdot Up_2\cdot \varepsilon^*)$, where
$\varepsilon_{\mu}(U_{\mu})$ is the polarization four--vector of the
$\rho (a_1-) $--meson and $q(p_2)$ is the four--momentum of the $\rho
(\pi -)$--meson. The expression for the width of this decay in the $a_1 $--
meson rest frame is
\ba
\Gamma (a_1\to\rho\pi )&=&\frac{m^2_{\rho}}{192\pi M^5}
[(M^2+m^2_{\rho}-m^2)^2-4m^2_{\rho}M^2]^{1/2}
(c_1|\bar f_1|^2+c_2|\bar f_2|^2+2c_{12}\mathrm{Re}\bar f_1\bar f_2^*),
\nn\\
c_1&=&8m^2_{\rho}M^2+(M^2+m^2_{\rho}-m^2)^2, \nn\\
c_2&=&3M^2(2m^2-M^2)+(m^2_{\rho}-m^2)^2+2\frac{M^2}{m^2_{\rho}}(M^2-m^2)^2, \nn\\
c_{12}&=&M^4+4M^2(M^2-m^2_{\rho}-m^2)-(m^2_{\rho}-m^2)^2. 
\label{eq:eqccc}
\ea

Let us calculate the square of the matrix element for the $a_1\to 3\pi $
decay using the helicity formalism. In the general case the $a_1$--meson is
described by the spin--density matrix and one can write:
\begin{equation}\label{8}
|M|^2=\rho_{\lambda\lambda'}B_{\lambda\lambda'}, \ \
B_{\lambda\lambda'}=A_{\lambda}A^*_{\lambda'},
\end{equation}
where $A_{\lambda}=U^{(\lambda)}_{\mu}A_{\mu} $ and $U^{(\lambda)}_{\mu}$ is
the $a_1$--meson wave function with definite helicity, $\rho_{\lambda\lambda'}$
is the $a_1$--meson spin--density matrix in the helicity representation. The
standard choice of the $z$ axis is along the normal to the decay plane in the
$a_1$--meson rest frame (we chose it along the direction ${\vec q}_2\times
{\vec q}_1$, where ${\vec q}_1({\vec q}_2)$ is the three--momentum of the
first (second) pion). Since all pion momenta are perpendicular to the $z$
axis the quantities $B_{\lambda\lambda'}$ with zero helicity $\lambda $ or
$\lambda' $ are equal to zero. Finally, one obtains:
\begin{equation}\label{9}
|M|^2=\rho_{++}B_{++}+\rho_{--}B_{--}+2\mathrm{Re}\rho_{+-}\mathrm{Re}B_{+-}-
2\mathrm{Im}\rho_{+-}\mathrm{Im}B_{+-}.
\end{equation}

As it is known, the decay of a particle into a three--body final state is
characterized by two independent variables. We choose the
energy of the negative pion $\omega_2$ and angle $\theta $ between the
momenta of the first and second pions, i.e., between ${\vec q}_1$ and
${\vec q}_2$. The $x$ axis is directed along ${\vec q}_2$. In this case
the quantities $B_{\lambda\lambda'}$ are
\ba
B_{\pm\pm}&=&\frac{1}{2}[{\vec q}_1^2|a|^2+{\vec q}_2^2|b|^2+
2q_1q_2(\cos\theta \mathrm{Re}ab^*\mp \sin\theta \mathrm{Im}ab^*)],
\nn\\
\mathrm{Re}B_{+-}&=&-\frac{1}{2}[{\vec q}_1^2|a|^2\cos2\theta +{\vec q}_2^2|b|^2+
2q_1q_2\cos\theta \mathrm{Re}ab^*], \nn\\
\mathrm{Im}B_{+-}&=&-q_1q_2\sin\theta \mathrm{Re}ab^*-\frac{1}{2}{\vec q}_1^2|a|^2
\sin2\theta , 
\label{eq:eq10b}
\ea
where $q_1 (q_2)$ is the value of the three--momentum ${\vec q}_1 (
{\vec q}_2)$.

The energy and angle distribution of the decaying $a_1$--meson is
described by the following expression
\begin{equation}\label{11}
\frac{d\Gamma (a_1\to3\pi )}{d\omega_2d\theta}=
\frac{1}{(2\pi )^4}\frac{q_1q_2}{4M}[M-\omega_2+\frac{q_2}{q_1}\omega_1
\cos\theta ]^{-1}|M|^2.
\end{equation}
The energy $\omega_1$ is not an independent variable and it can be determined
in terms of two independent variables using the following identity
$$2q_1q_2\cos\theta =M^2+2m^2-2M\omega_2+2\omega_1(\omega_2-M). $$

We introduced above the general spin--density matrix for the description of the
polarization state of the $a_1$--meson. In the coordinate representation its  expression is given by Eq. (\ref{eq:eq5}).

In the $a_1$--meson rest frame this formula is written as
\begin{equation}\label{12}
\rho_{ij}=\frac{1}{3}\delta_{ij}-\frac{i}{2}\varepsilon
_{ijk}s_k+3Q_{ij}, \ ij=x,y,z.
\end{equation}
This spin--density matrix can be written in the helicity representation
using the following relation
\begin{equation}\label{13}
\rho_{\lambda\lambda'}=\rho_{ij}U_i^{(\lambda )*}U_j^{(\lambda')}, \
\lambda ,\lambda'=+,-,0,
\end{equation}
where $U_i^{(\lambda )}$ are the $a_1$--meson spin functions which have the
$a_1$--meson spin projection $\lambda $ on the quantization axis ($z$--axis).
They are
\begin{equation}\label{14}
U^{(\pm )}=\mp \frac{1}{\sqrt{2}}(1,\pm i,0), \
U^{(0)}=(0,0,1).
\end{equation}
The elements of the spin--density matrix in the helicity representation
are related to the ones in the coordinate representation by such a way
\begin{equation}\label{15}
\rho _{\pm\pm}=\frac{1}{3}\pm \frac{1}{2}s_z-\frac{3}{2}Q_{zz}, \
\rho_{00}=\frac{1}{3}+3Q_{zz}, \
\rho_{+-}=-\frac{3}{2}(Q_{xx}-Q_{yy})+3iQ_{xy}, \
\end{equation}
$$\rho_{+0}=\frac{1}{2\sqrt{2}}(s_x-is_y)-
\frac{3}{\sqrt{2}}(Q_{xz}-iQ_{yz}),
\rho_{-0}=\frac{1}{2\sqrt{2}}(s_x+is_y)+
\frac{3}{\sqrt{2}}(Q_{xz}+iQ_{yz}), \ \rho_{\lambda\lambda'}=
(\rho_{\lambda'\lambda})^* .  \  $$
To obtain these relations the condition $Q_{xx}+Q_{yy}+Q_{zz}=0$ was applied.

The factor 1/3 in the unpolarized part of the spin--density
matrix was introduced, as for the decay it is necessary to average over the spins of the $a_1$--meson.

%%%%%%%%%%%%%%%%%%%%%%%%%%%%%%%%%%%%%%%
\section{Conclusion}
%%%%%%%%%%%%%%%%%%%%%%%%%%%%%%%%%%%%%%%

Using the parametrization of the electromagnetic transition $\gamma^*\to a_1(1260)\pi $ in terms of two FFs, we investigated the polarization phenomena in the annihilation reaction (\ref{eq:eq1}). We 
calculated  the differential (and total) cross section and various 
polarization observables as functions of the FFs. The spin--density matrix 
elements of the produced axial meson have been also calculated. Explicit formulae for the decay of the $a_1$--meson into three pions have been given.

FFs are complex in the time-like region and have been parametrized according to a VMD inspired $Q^2$ dependence, saturated by vector mesons. The parameters have been fitted to the data, assuming that all the four pion yield is due to reaction (\ref{eq:eq1}). If other intermediate channels contribute to this yield, and only a fraction of the cross section is due to the $a_1\pi$ intermediate state, then the normalization parameters should be rescaled by the square root of that fraction, assuming no dependence on $q^2$.

The $q^2$ dependence of the FFs shows, as expected, peaks in correspondence of the masses of the considered vector mesons. An interesting feature appears for $f_2$ at $Q^2\sim 4$ GeV$^2$, which is due to an interference of the different terms. Note that none of the considered vector mesons corresponds to a mass of such value. An interference of such origin can be the reason for the plateau seen in the cross section data, which, however does not appear with large intensity in our fit.

The reaction (\ref{eq:eq1}) has been clearly detected in the experiments. The present results can be useful for the analysis of the experimental data and for the determination of the $\gamma^*\to a_1(1260)\pi $ transition FFs. 

%%%%%%%%%%%%%%%%%%%%%%%%%%%%%%%%%
\section{Acknowledgment}
%%%%%%%%%%%%%%%%%%%%%%%%%%%%%%%%%

This work was inspired by enlightening discussions with Prof. M. P. Rekalo. Two  of us (G.I.G. and C.A.) acknowledge the hospitality of CEA, Saclay.  This work was partly supported by the Slovak Grant Agency for Sciences VEGA under Grant N. 2/7116/28 (CA) and  by grant INTAS Ref. Nr 05-1000008-8328 (G.I.G.). We acknowledge the french Groupement de Recherche Nucleon, for useful meetings and continuous support.

\end{document}